\documentclass[lettersize,journal]{IEEEtran}
\usepackage[includefoot,includehead,left=1.75cm,right=1.75cm,top=1cm,bottom=1.75cm]{geometry}
\usepackage{amsmath}
\usepackage{amsfonts}
\usepackage{graphicx}
\usepackage{enumerate}
\usepackage{booktabs}
\usepackage[absolute]{textpos}

\usepackage[hyphens]{url}
\usepackage{hyperref}
\hypersetup{pdfborder={0 0 0}}
\hypersetup{pdftitle={AI Ethics Principles in Practice: Perspectives of Designers and Developers}}
\hypersetup{pdfauthor={Conrad Sanderson, David Douglas, Qinghua Lu, Emma Schleiger, Jon Whittle, Justine Lacey, Glenn Newnham, Stefan Hajkowicz, Cathy Robinson, David Hansen}}
\hypersetup{pagebackref=false,breaklinks=true,colorlinks=true,bookmarks=true,urlcolor=blue,linkcolor=blue,citecolor=blue}

\usepackage[hang,flushmargin]{footmisc}   

\makeatletter
\def\footnoterule{\kern-3\p@
  \hrule \@width \columnwidth \kern 2.6\p@} 
\makeatother

\begin{document}

\title{AI Ethics Principles in Practice:\\Perspectives of Designers and Developers}

\author
  {%
  Conrad Sanderson, David Douglas, Qinghua Lu, Emma Schleiger, Jon Whittle,\\
  Justine Lacey, Glenn Newnham, Stefan Hajkowicz, Cathy Robinson, David Hansen\\
  \vspace{2ex}
  \textit{Commonwealth Scientific and Industrial Research Organisation (CSIRO), Australia}
  \vspace{-3ex}%
  }

\maketitle

\pagestyle{empty}
\thispagestyle{empty}

\begin{abstract}
As consensus across the various published AI ethics principles is approached,
a gap remains between high-level principles and practical techniques
that can be readily adopted to design and develop responsible AI systems.
We examine the practices and experiences of researchers and engineers
from Australia's national scientific research agency (CSIRO),
who are involved in designing and developing AI systems for many application areas.
Semi-structured interviews were used to examine how the practices of the participants relate to
and align with a set of high-level AI ethics principles proposed by the Australian Government.
The principles comprise:
(1)~privacy protection and security,
(2)~reliability and safety,
(3)~transparency and explainability,
(4)~fairness,
(5)~contestability,
(6)~accountability,
(7)~human-centred values,
(8)~human, social and environmental well-being.
Discussions on the gained insights from the interviews
include various tensions and trade-offs between the principles,
and provide suggestions for implementing each high-level principle.
We also present suggestions aiming to enhance associated support mechanisms.
\end{abstract}

\begin{IEEEkeywords}
AI~ethics, ethics principles, responsible~AI, responsible design, system design, artificial intelligence, machine learning, automation.
\end{IEEEkeywords}

\begin{textblock}{13.4}(1.3,14.9)
\hrule
\vspace{1ex}
\noindent
\footnotesize
\textbf{{$^\ast$}~Published in:} IEEE Transactions on Technology and Society, Vol.~4, No.~2, pp.~171--187, 2023. DOI:~\href{https://doi.org/10.1109/TTS.2023.3257303}{10.1109/TTS.2023.3257303}
\end{textblock}

\section{Introduction}

The rapidly increasing use of artificial intelligence (AI), which includes machine learning (ML),
throughout society has already produced significant benefits across a variety of fields,
and promises to deliver more in the future~\cite{Hajkowicz_2019,Koopman_2020,Liu_2021}.
As AI becomes more ubiquitous there are rising concerns regarding its use and development~\cite{Christoforaki_2022,Ryan_2021}.
AI may be used in applications that reinforce existing biases and historical disadvantage in society
(e.g., involving gender, race, socio-economic background),
infringe individual privacy, and make opaque decisions that affect people's lives (e.g., automated assessment of job applications)~\cite{Coeckelbergh_2020,Eubanks_2018,ONeil_2016}.
Responses to these concerns are well underway in many countries.
A~recent review of international principles and guidelines for AI ethics included 84 such documents~\cite{Jobin_2019}.
Out of the numerous ethics frameworks and guidelines produced by governments and organisations,
a~broad consensus is emerging around what the major principles of AI ethics should be~\cite{Fjeld_2020}.

Governments and organisations are beginning to develop mechanisms to make high-level AI ethics frameworks more actionable in practice.
Governance, regulation, and legal frameworks concerning the development and use of AI systems and associated technologies are emerging in many countries~\cite{Corinne_2018,Stahl_2021}.
The European Union has recently proposed a legal framework for AI for its member states~\cite{EC_2021}.
A~recent report by the Australian Human Rights Commission has recommended introducing legal accountability for both government and public sector uses of AI,
as well as establishing an independent AI Safety Commissioner~\cite{Farthing_2021}.

However, there is still a large gap between high-level AI principles and practical techniques
that can be readily used in the design and development of responsible AI systems~\cite{Mittelstadt_2019,Morley_2021a}.
More specifically, current high-level principles appear more focused on end users and people affected by AI technologies,
rather than developers of AI systems.
Furthermore, ethics principles are just one of a range of governance mechanisms
and policy tools necessary to promote the ethical development and use of AI technologies.

In this work semi-structured interviews are used to better understand
the existing practices and perspectives of scientists and engineers 
that develop and/or use AI/ML technologies across a broad range of projects
and application domains.
The insights and challenges gleaned from the interviews 
are analysed within a wider context, 
aiming to provide suggestions and caveats which may be helpful
when placing the high-level AI ethics principles into practice.

The interview participants are from 
the Commonwealth Scientific and Industrial Research Organisation (CSIRO),
which is Australia's national science agency.
The organisation is responsible for conducting scientific research in many industrial application areas,
including biosecurity, health, agriculture, and environment.
Recent efforts include increasing the uptake of AI/ML across these areas.
The Australian Government's voluntary high-level AI ethics principles~\cite{DISER_2020}
are used as a framing structure for the interviews,
as well as the subsequent analysis.
This set of principles is treated
as a representative of the many similar principles proposed by other governments and organisations~\cite{Fjeld_2020,Jobin_2019}.

We continue the paper as follows.
Section~\ref{sec:related_work} overviews related work.
Section~\ref{sec:methods} describes the methods used to conduct the interviews and the subsequent thematic analyses.
Section~\ref{sec:results} presents the salient results from the interviews,
in terms of observations, practices and challenges for each of the high-level ethics principles.
Section~\ref{sec:discussion} provides a discussion of the insights from the interviews
and provides suggestions for implementing each high-level principle.
Section~\ref{sec:recommendations} provides suggestions to enhance associated support mechanisms
at organisational and design levels.
Concluding statements are given in Section~\ref{sec:conclusion}.

\section{Related Work}
\label{sec:related_work}

The development and deployment of responsible AI requires users to adopt
a largely implicit set of high-level ethics principles into explicit practices.
High-level principles such as transparency and explainability
may require various approaches to implement within an AI system depending on its purpose 
and the user requirements~\cite{Arrieta_2020}.
However, the availability of useful ethical design specifications
for AI designers and developers is currently lacking~\cite{Morley_2021a},
with a recent review highlighting various limitations~\cite{Morley_2020}.
These limitations include the lack of methods that support the proactive design
of transparent, explainable and accountable systems,
in contrast to the less useful post-hoc explanations of system outputs.
Existing methods overemphasised AI `explicability'
(which is necessary but not sufficient for transparency and explainability),
and skewed to assessing the impact of an AI system on individuals rather than on society or groups.
Finally, the examined methods were found to be difficult to implement and typically positioned as discourse aids to document design decisions~\cite{Morley_2020}.

The limited support for tools available for designers and developers
demonstrates a gap in what is available and what is useful
for adopting responsible, or ethical, design approaches~\cite{Mikalef_2022,Morley_2021a}.
The implementation of ethics principles in practice
requires an improved understanding of the practices of designers and developers of AI systems,
and how they relate to high-level ethics principles.
By building this understanding, organisations can better position themselves
to develop the support required to produce and use AI systems responsibly. 

A recent interview study involving high-level organisational leaders (such as CEOs)
provides empirical evidence that a promising approach to operationalise AI ethics principles
is via four sets of practices~\cite{Seppala_2021b,Seppala_2021}:
(i)~governance,
(ii)~{AI} design and development,
(iii)~competence and knowledge development,
(iv)~stakeholder communication.

The operationalisation of AI ethics can be also framed as a design process
rather than an end goal \cite{dAquin_2018,Morley_2021b}.
Fostering an understanding of the complexities and challenges of implementing responsible AI
provides the foundation for iterative and adaptive approaches
rather than as the completion of activity checklists.
The implementation of responsible, or ethical, design has been characterised across
three complimentary design approaches~\cite{Dignum_2019}, summarised below.

~

\begin{enumerate}[{$\bullet$}]

\item
\textbf{Ethics \textit{in} Design}, 
which requires designers and developers to consider the purpose of the 
AI system under development, and the likely consequences of its use~\cite{Dignum_2019}.
Part of this involves considering the values (or motivating principles) reflected in 
the design and purpose of the AI system~\cite{Ferrario_2017}. This has parallels with efforts to 
incorporate human values in software engineering~\cite{Whittle_2019,Whittle_2021}. Anticipating 
the negative effects of failing to account for an ethical principle may also highlight 
potential issues that might otherwise be overlooked~\cite{Mikalef_2022,Agerfalk_2022}.

\item
\textbf{Ethics \textit{by} Design},
which concerns the behaviour of AI systems in deployment. The functions of systems 
should reflect ethical principles, such as minimising harm. This also covers 
constraints imposed on permissible actions and decisions the AI system may make 
\cite{Dignum_2019}. This is particularly significant for autonomous systems that may come into 
contact with humans or animals during their operation, and for AI systems that 
may be used to recommend decisions that affect people's lives.

\item
\textbf{Ethics \textit{for} Designers},
which  refers to the ethics that motivate and govern the developers of AI systems, and 
may include professional codes of conduct, ethical principles, and regulatory 
requirements~\cite{Dignum_2019}. These design approaches characterise the complex and dynamic 
task of developing responsible AI. In turn, the infrastructure required from 
organisations and governing bodies to effectively support responsible AI must be 
responsive to those complexities and the evolving nature of the field.

\end{enumerate}

\begin{table*}[!htb]
\caption
  {%
  \rm\normalsize
  An adapted summary of the voluntary high-level ethics principles for AI,
  as promulgated by the Australian Government~\cite{DISER_2020}.
  }
\label{tab:principles_summary}
\small
\centering
\renewcommand{\arraystretch}{1.2}
\begin{tabular}{p{0.35\linewidth}  p{0.55\linewidth}}
\toprule
\textbf{Principle} & \textbf{Summary} \\
\midrule
Privacy Protection and Security & AI systems should respect and uphold privacy rights and data protection, and ensure the security of data.\\
\midrule
Reliability and Safety &  AI systems should reliably operate in the context of their intended purpose throughout their lifecycle.\\
\midrule
Transparency and Explainability & 
There should be transparency and responsible disclosure to ensure people know when they are being significantly impacted by an AI system, and can find out when an AI system is engaging with them.
Explainability includes what the AI system is doing and why, and may include the system's processes and input data.\\
\midrule
Fairness &
AI systems should be inclusive and accessible, and should not involve or result in unfair discrimination against individuals or groups.\\
\midrule
Contestability &
When an AI system significantly impacts a person, group or environment, there should be a timely process to allow people to challenge the use or output of the system.\\
\midrule
Accountability &
Those responsible for the various phases of the AI system lifecycle should be identifiable and accountable for the outcomes of the system, and human oversight of AI systems should be enabled.\\
\midrule
Human-centred Values &
AI systems should respect human rights, diversity, and the autonomy of individuals.\\
\midrule
Human, Social and Environmental (HSE) Well-being &
AI systems should benefit individuals, society, and the environment.\\
\bottomrule
\end{tabular}
\vspace{1ex}
\end{table*}

\section{Methods}
\label{sec:methods}

The interview participants were
CSIRO scientists and engineers  who design, develop, or implement systems with AI components (such as ML).
An initial set of participants was sought via calls for participation circulated within the organisation,
where the main selection criterion was self-reported involvement with research 
and/or development work that notably included the use of AI methods and/or technologies.
The initial phase was followed by a snowballing technique where additional participants
were recruited through suggestions by preceding participants.
The number of participants increased until a saturation point of perspectives
was deemed to have been reached~\cite{Colvin_2016,Hall_2015}.
In total, 21 interviews were conducted.

The participants represented a cross-section of experience and responsibility;
the positions of the interviewees included:
team leader~(8), 
group leader~(4),
principal research scientist~(2), 
principal research engineer~(1), 
senior research scientist~(4), 
research scientist~(1), 
postgraduate student~(1).
The gender division was approximately 24\%~female and 76\%~male. 

Each participant was provided with a summarised version
of the Australian AI ethics principles (see Table~\ref{tab:principles_summary}),
prior to their interview.
The interview protocol aimed to first evoke
3 to 4 principles that were most pertinent to each participant
as encountered in their work.
Each of the principles chosen by the participant
was then examined through the following prompts:
\textbf{(i)}~how did the principle express itself in their work,
\textbf{(ii)}~how was the principle addressed,
\textbf{(iii)}~what tools/processes would be useful for addressing the principle.
Subsequent questions aimed to cover relevant overlaps with the following areas:
software development, machine learning, and ethics in AI.
Participants were then asked to consider other ethical factors
not covered by the high-level principles but encountered and potentially addressed in their work.

The interviews were conducted in a conversational setting
by three interviewers that had significantly different research backgrounds and experience.
The interviews ranged in length from 22 to 59 minutes, 
with a median length of 37 minutes.
All the interviews were done from February to April 2021.

\subsection{Thematic Analysis}

The transcripts of the interviews were analysed using methods from the `codebook' school of thematic analysis,
where themes are understood as `domain summaries' and given before starting analysis,
and the codes developed during the analysis are assigned to these themes~\cite{Braun_2019}. 
The high-level principles (see Table~\ref{tab:principles_summary}) were used as the domain summaries.
Concepts in the text were given a broad theme and a more specific subtheme (if necessary),
and were attached to the most relevant high-level principle.

The analysis was accomplished at a semantic level,
which focused on describing and interpreting patterns identified in the interview transcripts,
instead of looking for underlying assumptions or concepts in the transcripts~\cite{Braun_2006}.
The thematic analyses derived from the interview transcripts were independently cross-checked 
by the three interviewers to ensure consensus on themes and inter-rater reliability.

\subsection{Ethics Approval and Consent}

This study was approved by CSIRO's Social and Interdisciplinary Science Human Research Ethics Committee,
in accordance with the National Statement on Ethical Conduct in Human Research~\cite{NMHRC_2018}.
The approval included the scope of the study, the selection and contact of potential participants,
the interview protocol, and the de-identification requirements for the transcripts.
Informed consent was sought and obtained from each participant prior to their interview.

\subsection{Limitations}

The interview participants were selected via solicitation emails and peer suggestions,
which may pose a threat to internal validity.
The selection and availability of interviewees may have constrained
the span and nature of perspectives and hence outcomes of this study.
Although selection bias is possible when the participants are not randomly selected,
the threat to internal validity is partially alleviated
as the interviewers had no contact with the interviewees beforehand.
To reduce the risk of missing information and interviewer subjectivity,
each interview included three interviewers
with significant differences in their research backgrounds and experience.
The interviewers worked jointly to pose questions,
which aids in increasing the depth and range of inquiry,
as well as reducing the likelihood of subjective bias.

While this study aims to include Indigenous viewpoints,
no information was sought about the interviewees' ethnic background
and none of the interviewees provided such information.
This may have affected the interpretation and expression of Indigenous viewpoints.

This study was undertaken in one organisation,
which may pose a threat to external validity;
the opinions provided by the interviewees may not be representative
of the larger AI development community.
To reduce this threat, we ensured that the interviewees had various roles and degrees of expertise,
and worked on a variety of research areas and projects (for both internal and external customers).
While having interviewees from more organisations would be desirable,
we believe our study can be relevant to many AI/ML researchers and system development teams;
the observations and comments given by our interviewees
provide insights into the challenges developers are facing in dealing with AI ethics issues
during research, development and deployment.

\section{Results}
\label{sec:results}

In this section we present the salient points resulting from the thematic analysis of the interviews,
covering practices, challenges and insights relating to each of the high-level principles shown in Table~\ref{tab:principles_summary}.
The occurrence of themes related to the principles
across the interviews is shown in Table \ref{tab:occurrence}.

Quotations and paraphrased sentences are ascribed to interview participants 
via markers in the shorthand form of P{\#\#},
where {\#\#} is a two-digit participant identifier (01 to 21).

\begin{table}[!b]
\vspace{1ex}
\small
\centering
\caption{\rm\normalsize Occurrence of themes related to AI ethics principles within the 21 interviews.}
\label{tab:occurrence}
\begin{tabular}{p{0.65\columnwidth}p{0.25\columnwidth}}
\toprule
\bf{Principle} &
\bf{Occurrence}  \\
\midrule

Privacy Protection and Security & 17 / 21  (81\%) \\
\cmidrule{1-2}

Reliability and Safety & 19 / 21  (90\%) \\
\cmidrule{1-2}

Transparency and Explainability & 18 / 21  (86\%) \\
\cmidrule{1-2}

Fairness & 10 / 21  (48\%) \\
\cmidrule{1-2}

Contestability & ~8 / 21  (38\%) \\
\cmidrule{1-2}

Accountability & 13 / 21  (62\%) \\
\cmidrule{1-2}

Human-Centred Values & ~3 / 21  (14\%) \\
\cmidrule{1-2}

HSE Well-being & 11 / 21  (52\%) \\

\bottomrule
\end{tabular}
\end{table}

\subsection{Privacy Protection and Security}

For addressing this principle,
participants suggested methods such as federated machine learning~\cite{TianLi_2020,QiangYang_2019},
aggregation of data,
removing personally identifiable information,
and differential privacy~\cite{Dwork_2008,Friedman_2010}.
As an example, partitioning data across distributed machines was mentioned as a method of 
protecting privacy for genomic data:
``[various] research institutions from around the world can collaborate,
because they don't have to give up their data.
They don't have to share their data in the clear''~(P03).
There are also strong practical reasons for doing so:
``[t]he machine learning methods that are running on those,
they are split across multiple nodes anyway,
[...] with the idea that they only ever use a portion of the genome.
Not because it's safer, but because it's more efficient.
It's faster [...], that's the reason why it's possible to analyse the whole genome at the same time''~(P03).

The impact on ML models trained using altered data
(with personally identifiable information removed) is uncertain~(P19). 
Differential privacy may also have a negative impact of the utility of ML models~(P05).
The concerns about the impact of differential privacy methods and removing PII 
are examples of the tension between protecting privacy and producing accurate ML models:
``that's the real challenge that we are always dealing with, this privacy and utility trade-off''~(P05).
There are also possible trade-offs between privacy and other principles,
such as \textit{Reliability}, \textit{Accountability}, as well as \textit{Transparency and Explainability} (P01, P07, P09).
For example, to determine the accountability for failure in a process may involve using
sensitive information that stakeholders are not willing to share due to their stated requirement for privacy~(P07).

Privacy concerns were mentioned as a reason to avoid using sensitive data:
``we have deliberately worked on data that doesn't have any privacy concerns,
because of the difficulty of working with data that does''~(P09).
Synthetic data and standard machine learning datasets were mentioned as alternatives to using sensitive data~(P05).
Minimising the amount of data needed was also mentioned as a design goal~(P02).

Gaining and maintaining the trust of collaborators and data providers
was identified as an important motivation for privacy and data protection.
``Whether their concern is correct or not, that doesn't even matter here.
If I don't get that trust, they're not collaborating''~(P07).
Similarly, another participant observed that in their practice
``you build the trust with the data providers [...]
so more people can give you data and increase your data representability''~(P02).

Approval from the human research ethics committee was mentioned as important for 
confirming the requirements for working with sensitive data~(P11, P17, P19).
Privacy impact assessments were also mentioned as another process for identifying potential privacy issues~(P10), 
in order to understand privacy concerns, build provisions to enable privacy controls,
and allow people to highlight things that they did not want to make visible in the platform.
These assessments may highlight issues that require consultation with external clients:
``we've had to manage [...] our client's expectations
around [the] additional effort needed in the privacy [...] side of things''~(P10).

External collaborators may impose access restrictions on their data~(P01, P17).
Clients may also require that data remains within their network environment,
which requires researchers to perform their work on the client's servers~(P17, P19).
The access restrictions attached to sensitive data may prevent projects from proceeding:
``sometimes early meetings with potential customers [...],
we'll raise that quite early as saying, what data do you have and what's the access like to that data?
Because sometimes that's the gatekeeper for continuing the project or not''~(P17).
Concerns were raised about provision of data to third parties without explicit agreement:
``I gave you data and I consent for a particular application and that's it.
If somebody else wants my data they should ask me''~(P11).

One interviewee suggested that access to sensitive data can be partial and conditional:
``[y]ou can [allow] certain applications to access certain areas of your genome.
So if it's a research project [...] you might enable them to see the whole data at the same time''
(P03).
Furthermore, in the context of genomics, the data would be encrypted
and ``you as the patient own the [decryption] key''~(P03).

Privacy concerns differ from project to project~(P10). 
Accidental collection of sensitive data was mentioned as an issue for autonomous systems~(P12, P13).
Projects that involve Indigenous data need to respect the cultural context of that data:
``[s]o when we talk to Indigenous people about where data is and who owns it and who can benefit from it,
it gets affected by things like whose country it is,
whose knowledge is going to be part of that, is it gendered spaces,
is it sacred spaces that we are looking at?''~(P14).

\subsection{Reliability and Safety}

Participants often mentioned that the quality and amount of training data is very important.
One interview noted that ``if you're training a model with not a lot of data
and you come across an edge case, you can actually get some really weird results''~(P09).
Obtaining a sufficient number of samples can be challenging,
as obtaining one sample can be high in terms of both financial cost and time cost in domains such as genomics~(P03).
Furthermore, awareness of potential mismatches between training data and data used during deployment is necessary
to prevent the trained model from being unsuitable for its intended purpose~(P04, P18).
During system development, it is also important to consider how to respond to incomplete data:
``[D]o we just operate in its absence or do we assume it's true or assume it's false
or make some other determination?''~(P16).
Hidden variables that aren't captured within the data are a significant problem~(P02).

Prior information can be used to reduce the need for large amounts of training data.
``So, if you have a weather model, climate model, and some chemistry process,
built inside, then you probably need the less of this data''~(P02).
However, using prior information can be difficult within machine learning models.
``[Y]ou can say it's a prior model, [...] but it's drawn from [...] scientific literature,
or based on the expert knowledge, but it's really hard to see that [it's]
some nice distribution that you can easily put into the model''~(P02).

One interviewee pointed out that there might be multiple AI algorithms suitable for a task,
and developers need to assess potential candidates according to several categories~(P09). 
Instead of focusing only on accuracy, 
there are ``trade-offs in terms of data complexity versus accuracy''~(P09).
Furthermore, robustness can be adversely affected:
``the push for accuracy is not always beneficial,
because it can include complexity and perturbation effects,
[...] and then you have these situations that basically explode because of their complexity''~(P09).
Within the field of autonomous systems, it was noted that 
although machine learning is a trend and works well for many types of applications (e.g., recognition),
it is important to find the right mix of traditional proven technology
and newer machine learning technology~(P12).

Bugs in the implementation of AI systems can produce
``erroneous results, erroneous inferences, [and] could have led to dangerous decisions downstream''~(P16).
Failing safely is deemed to be of prime importance and is considered ``as you're building the system''~(P16).
However, ``there's only so much you can think ahead, about what those failure modes might be''~(P16).

As degradation of system performance may occur during deployment,
AI systems may need recalibrated on new data.
``If you build a model on 10 year old data,
then you're not representing the current state of risks for certain disease.
[...] as a minimum, [recalibration] on new data would probably be [...] more meaningful''~(P18).
Furthermore, the systems may need to be completely rebuilt to properly take advantage of newer and/or more comprehensive data which was not considered during the initial deployment~(P18).

AI systems were also noted as being difficult to verify:
``you can always have a pure statistics approach and build the system and let it run in the environment.
But when do you have a complete assessment really?
Especially with systems that change over time and based on sensory input
[...] it's very difficult''~(P12).
How AI models may be verified will depend on their context. 
Where ML models are being used to emulate existing processes, 
expert judgement is necessary to determine how close the model's predictions are 
to the modelled process~(P04).

The element of safety also extends to the physical context within which an AI system will be used.
As one participant explained, ``it's not just the AI data that has to be safe,
it's actually its application and use [...] reliability and safety, throughout its whole life cycle
and its application into the real world become really critical questions''~(P14).
Experts in all associated disciplines
(hardware engineers, software engineers, end users, and other stakeholders)
should be involved in the design~(P12).
AI developers rely on domain experts to ascertain whether the AI system is
correctly following existing legal rules in the application domain~(P06).

Tracking the use of an AI system has been suggested as useful for building user trust in the reliability
of the system~(P15).
Furthermore, in conjunction with the ability to override the system's suggestions,
usage tracking would allow users to note the differences in outcomes:
``[the system] suggested doing one scenario, we chose to do another, this is the result we got [...] 
did we do the job that we expected? Or did we do the job that the system expected?''~(P15).

The need for large amounts of data to improve accuracy is in tension
with the goal of minimising data collection and use to protect privacy~(P02). 
In tension with the \textit{Fairness} principle, 
in certain scenarios (such as medical applications),
developers may have to restrict the scope of data as much as possible 
(e.g., use data only from people within an ethnic group),
in order to make it easier to build an AI solution that works as intended~(P11).

A concern raised by interviewees is the usage of AI systems after their development (i.e.,~during deployment).
One noted an instance where their AI system was designed to provide recommendations
according to criteria agreed with a client,
and yet these recommendations were ignored in decision making by the client
as other criteria were used in practice~(P08).
Another participant noted that nuances about the collected data and the assumptions made in
analysing the data used to train an ML model are likely to be lost when the model is used in practice:
``you can put all those cautions in place but they won't necessarily be respected in practice
and you've got very little control over that''~(P01).
Due to the lack of control over the use of AI technologies after their development,
some researchers and/or developers may disclaim accountability:
``whether the [system] works in 10 years, it's not under our control [...]
and we shouldn't really care about it''~(P11).

In the context of autonomous systems, 
the operations of drones are regulated
but AI technology embedded in drones is currently not regulated~(P12).
The regulations include that ``autonomous flight is prohibited [...]
in the sense that the flight is not managed by [a remote] human pilot''~(P12).
However, this raises a question of reliability:
``if [there is] a [target following] mode enabled in [the] system [...]
you have to have a human pilot monitoring this system and disengaging that system at any time
[...] so now the question is [...] 
what happens if [the] remote pilot is really there,
flicks the [disable] switch and the system doesn't react?
The remote pilot is not always in full control of [the UAV]
because of technical reasons''~(P12).

Reliability and safety are significant concerns for autonomous systems:
``how dependable is the autonomy?''~(P12).
Where AI is used in robotic systems, the quality of the hardware used in the system
is an important safety consideration~(P12).
The hardware underneath needs to be certified or qualified to industry standards~(P12).

\subsection{Transparency and Explainability}

In discussing this principle, the majority of participants focused on the explainability of decisions made by AI systems.
One participant observed: 
``Whenever you provide estimates or outputs from an algorithm,
they need to be explained in some way,
and so you need to be very open and transparent about what you've done,
how you've pre-processed your data, 
how you've applied the algorithms,
even some of the fine-tuning of the algorithms [and] the hyperparameters''~(P09). 
Transparency and explainability were also seen as important for accountability to clients and users:
``[...] it's essential, because when you give a prediction, or when the prediction is wrong [...],
then you have to give people some sort of explanation''~(P02).

The degree of possible explainability depends on the form of AI used.
Symbolic AI uses defined rules to process data,
and allows for the chain of reasoning used to produce an output to be identified.
In contrast, statistical (or non-symbolic) AI, such as that used in machine learning,
develops rules from the data used to train it~\cite{Koopman_2020,Bishop_2006}.
Statistical AI methods were generally thought to be less transparent and explainable than symbolic AI.
One participant noted: ``the AI I use are equations -- we can't solve these equations easily;
we develop an algorithm for solving these equations, but we know what's happening.
We know what these algorithms are doing. It's not a black box.
The issue is the solution then is quite complex to understand
so that's where explainability of a solution and explainability of AI is a very important topic for us''~(P08).

The factors and data used to make a particular recommendation may also be presented to the user:
``it gives them a way to [...] trace back to the actual patient record [...]
to see why the system thought what it did''~(P19).
The level of explainability may also depend on the intended audience~(P17).

Explainability may weigh more than performance regarding trust in AI.
Fine-grained interpretability is needed in AI systems to be clear
which part of the outcomes are substantiated and which parts are uncertain~(P03).
Showing the current step in the decision-making process is helpful to improve user trust in ML models~(P11).
The output of the model is often not that useful to assist users to make decisions,
unless the system shows the indicators and factors for why that prediction was given~(P17).
One participant pointed out when the team had initial stakeholder meeting,
no one asked about the predictive performance of the algorithm,
but wanted to have a look at risk profiles behind the risk score~(P18). 

Participants also noted that trade-offs between accuracy and explainability may be necessary.
There are algorithms which provide good results but cannot explain how they arrived at that results~(P19).
``You may end up choosing possibly a suboptimal method that is transparent and explainable,
as opposed to the optimal method that [...] is [a] black box''~(P17).
Another participant stated that
``there have been instances where we've chosen an explainable model
which has slightly [lower] performance [than] a non-explainable model
which has higher performance but would be harder to convey the reasoning behind the prediction''~(P18).

Transparency was often linked to the norms of scientific publications
and access to scientific data to allow reproducibility of results.
``Transparency has to do with an open science, you want people to be able to reproduce your results''~(P11). Releasing the source code for AI applications was also mentioned as a means of transparency,
especially as part of publishing research~(P11).
However, practical concerns often prevent this from occurring:
``often it doesn't happen because it costs money to do,
you have to spend time to clean it up, to maintain it, to publish it and so on.
Second, you decrease the commercial value of it usually''~(P11).
Another participant noted that it would be useful to have official guidance
about the requirements for publishing AI source code as open source as part of research publications~(P09).

One hurdle for transparency is the use of proprietary data in the training of AI models.
The use of non-releasable proprietary data makes reproducibility difficult
for researchers outside of the original project team~(P13).
Clients may also require any research publications about an AI application
to be approved by them before submission~(P09).
This is part of a broader tension between openness and commercial incentives:
``Where it becomes a little bit challenging is where there's commercial in confidence [information]
or the commercial sensitivities around what you're doing,
and you may not be wanting to disclose exactly what you've done''~(P09).
This tension may be resolved in initial client negotiations:
``[our clients] understand we are a research organisation and one of our KPIs is publications as well.
So this was all put on the table and discussed in the negotiation phase''~(P13).

One response to making AI research reproducible is to use public datasets~(P11).
A further alternative was to use synthetic data:
``you can have synthetic data sets that carry the [...] the same sort of statistical properties 
so that you can then run your algorithms and so you can make them open source 
and [...] people can test algorithms and do a bunch of independent review and checking 
[...] without using the actual data''~(P01).

It was also noted that clients needed to be transparent in their claims
of how AI applications are (or are not) being used: 
``we want to make sure we deliver the tool and then it's up to the stakeholder decision maker
to follow or not follow our recommendations - but I want them to not misrepresent
the way they're making their recommendation''~(P08).

One participant argued that an understanding of risk was more important than explainability:
``[o]nce I know that it works most of the time I don't need explainability, I don't need transparency.
It's just temporary to establish the risk profile''~(P11).
Another participant noted that transparency is only a concern if something goes wrong,
and drew a parallel with relying on air travel without understanding how aircraft operate~(P02).

Traditionally, in statistics, people have used simpler linear models and assign meaning to the parameters.
In machine learning, developers often only think around the predictions~(P21).
The meanings connected to parameters are less well understood
in machine learning techniques as opposed to more classical statistical techniques
although they are really the same thing in a certain sense~(P21).

\subsection{Fairness}

Participants interpreted fairness in various ways.
P03 understood fairness as training ML models on diverse data,
and discussed several methods of addressing bias within the data used to train the AI system.
Omitting data from over-represented groups so that groups are more equally represented in the data was one method,
although there is the risk that the data removed will reflect the researchers' own cognitive biases~(P03). 

Downsampling populations that are over-represented in the data was another method of addressing bias,
although it was described as
``a very crude approach [to] the problem because you're getting rid of samples that you think are going to be identical. If they are then that's probably fine.
But in the vast majority of the complex applications that we have, there is probably still a subtle variation in there that might be the crucial thing for the machine learning methods to pick up. You don't know from the beginning''~(P03).
Placing judiciously selected weights on data was suggested
as a method of maintaining variation in the data
while also countering over-representation in the data~(P03).
However, manually adjusting the weights has its own problems:
``you're going in with an assumption that you know how to tune this thing,
which basically goes against machine learning as such, [...] where it's purely data driven''~(P03).
Even though mitigation approaches may address some problems, biases in the data were still seen as an inevitability~(P03, P09).

Incomplete data was identified as a problem for ensuring fairness.
``Sometimes you can be limited in what data [you have] available to use in the first place''~(P01).
As one participant noted, ``you can only tweak the data that you have''~(P03).
The cost of obtaining data was also noted as a problem~(P03).

Due to the potential impact of biases within the training data affecting the recommendations of the AI system,
user judgement in accepting results (or a `human-in-the-loop') was suggested as a potential response~(P03).
User judgement may be supported by indicating whether the sample being tested
has a close resemblance to the data used to train the ML model~(P03).
Another participant noted that it is important for users to understand the context of the system
so that the representations of people made by that system are judged fairly~(P10).

For historical clinical data, discriminatory language was noted as a potential problem for fairness~(P16).
``[A]ll the clinical concepts [...] have a preferred label associated with them and a set of synonyms.
Of course, over time, the social acceptability of certain terms changes.
So, we've provided feedback to ensure that, where we've come across those terms,
that they get updated in the reference data''~(P16).

Some participants noted the differences in interpreting fairness as `non-discrimination'
and as the opposite to `unfair discrimination'.
The removal of identifying information from data was mentioned as a method of ensuring fairness~(P07).
This is fairness as non-discrimination.
However, discrimination (in the form of exclusion criteria)
was mentioned as a necessity for research using medical data:
``[y]ou want to remove as much variance as possible to find out the signal that you want,
and it comes only later to see if that signal is applicable to other groups''~(P11).
This discrimination would not necessarily be `unfair discrimination'.
The same participant also noted that
``there are applications where you absolutely want it to be fair for everybody the same way.
For example if you do a loan application, a loan application if you have an [AI system],
you know that for this particular one it has to be the same no matter what is the population''~(P11)

Fairness in the distribution of benefits and burdens from using the AI system was also mentioned:
``the people [bearing the] costs should be the ones also getting the benefits out of it in the end''~(P07).
The benefits of AI systems trained on biased data may be unevenly distributed~(P02).

Another participant questioned whether fairness is necessarily linked with transparency:
``[w]hat's interesting about that fairness principle is it assumes that openness and transparency is going to be our only pathway for an ethical outcome''~(P14).
Maintaining the ownership of data may conflict with the goal of making the data accessible to others~(P14).
However, while the data used by the AI system may not be open and accessible to others,
the methods developed to analyse and present the data can be made open and accessible~(P14).

The role of fairness was unclear in contexts where the AI system is not used to make decisions
that directly impact individuals, communities, or groups~(P08).

\subsection{Contestability}

Where the AI system is used for decision support, contestability is implemented 
by allowing the user to overrule the recommended decision~(P18). This was 
explained with an example of a system used in clinical contexts: ``there was 
actually a defined process where if a patient was not flagged as being high risk, 
[...] clinicians were still allowed to include the patient into the next step 
clinical review''~(P18).  

One issue identified with contestability is the difficulty of revising the initial 
assumptions and design choices made in collecting and analysing the data used by 
the AI system:  ``if your results or the product is somehow sensitive to initial 
assumptions or choices made [...], it's not like you're going to be able [...] to 
run things again with different assumptions and choices''~(P01).

Another problem for contestability is the complexity of the explanation of the 
AI system's decision making~(P03). This complexity may prevent 
clients and users from contesting the system's output: ``it's just another layer 
of complexity that I don't think they'll be interested in or in a position to 
understand it. But I don't know whether they would be able to critically 
evaluate or contest, as you say, what has come out of it''~(P20). 
Another aspect of this was data where the relationships were not well 
understood. Genomic data was given as an example: ``basically, there is no 
one-on-one relationship between typically higher blood counts, say, and these 
other negative outcomes. For genomic [data] it could be the whole range. So 
therefore this interpretability is really not as clear-cut''~(P03).

Contestability was also interpreted as interpretability of an AI system's 
decisions, and overlapping with the principle of transparency and explainability~(P03).
One example mentioned was the logging and time stamping data collected by autonomous systems:
``if someone says [...] you took data [...] there where you were not allowed to do this and whatever.
I mean we can say this is what happened. So it's all logged''~(P12).

One interviewee pointed out it is hard to get people to trust an AI system that 
is just telling them to do something but doesn't give them the choice to 
disagree with the system~(P15).

\subsection{Accountability}

Transparency and explainability were seen as important for accountability: ``if you 
want some accountability, you need to find [an] explanation. So, the model [needs] 
to support the interpretability''~(P02). Transparency assists accountability 
by allowing domain experts to identify anomalies in the data~(P18). The 
reproducibility of published results is an important form of accountability for 
academic researchers. ``For us, the only bad thing is we publish a paper and then 
people try to reproduce it, and nobody can''~(P11).

Testing the AI system was identified as important for accountability. Test data 
may be developed in collaboration with the client to evaluate the system~(P18).
The system may also be evaluated empirically: ``that'll be done in terms of having 
some unseen set of data, which is human labelled, has some ground truth, and that 
we evaluate the system on that''~(P17). Accountability for the data used 
to train the ML model was attributed to the provider~(P16).

There were differences in opinion over accountability for the outcomes of decisions 
based on an AI system's recommendations. One researcher noted that ``the creator of 
technology is as responsible for what happens to the technology as the people who 
make the decision how to use the technology''~(P12). Similarly, another 
observed that ``at some stage, your prediction is wrong and there's some loss, then 
who is responsible for this? I believe the problem exists now, but, when you deploy 
a model, the problem can be more complicated. [...] [S]ome sort of framework should 
be established for who actually responsible for this''~(P02). However, 
another argued that this form of accountability did not apply to projects that are 
still within the research and development stage~(P13). The lack of control 
over the use of AI technologies after their development was also noted 
as a justification for disclaiming accountability by developers:
``Whether the [system] works in 10 years, it's not under our control''~(P11).

Accountability is also important for ensuring the reliability and safety of AI systems.
Logging and recording of metadata used by the AI system in making decisions is important for reviewing the system's output~(P04).
Accountability is important for the adoption of autonomous systems:
``I think with ML, we've seen examples [...], especially in the autonomous car industry,
where there's research being done into accidents and [...] if you can dig into detail what happened,
why that happened, I think that brings a lot more security and safety,
[...] both commercially and to the customers as well''~(P13).

For Indigenous end-users and collaborators, accountability extended beyond the outputs of the AI system
to the larger process of applying an AI system itself:
``The other part of that accountability though was that [Indigenous collaborators and co-creators]
said we have to benefit from the process, not just the outcome''~(P14). 
AI projects were accountable to indigenous collaborators and co-creators
for the capacity-building with digital technologies that the projects enabled with their indigenous partners~(P14).

\subsection{Human-Centred Values}

Participants noted the difficulty in clearly noting the presence of human-centred values in their projects:
``Values are often implicit. They're not stated explicitly in things'' (P1).
It was also observed that human-centred values were less applicable to applications
that did not involve direct human contact with the AI system~(P12).

Human-centred values were mentioned largely in relation to projects
co-created with Indigenous groups that are built on Indigenous knowledge.
The cross-cultural context where the AI system is used needs to considered:
``The problem is that sometimes when we design AI,
we don't take that care to think about which system it's going to be put into.
We just do it as a prototype to then to be deployed everywhere''~(P14).

Indigenous knowledge, rights and practices should also be incorporated
into the design and deployment of AI systems,
including how and why data is collected where data will be physically stored~(P14).
AI systems for Indigenous groups should also reflect their ethics and  worldviews:
``I think it was really interesting to actually push that a bit further
and say actually, let's make really explicit biases in [an Indigenous-focused project]
that actually [make] sense to Indigenous people.
So we didn't just do it on (...) human-centred [values],
we really pushed it to say how do we make [the AI] learn
so that it actually is really biased to an Indigenous group''~(P14).

\subsection{Human, Social and Environmental (HSE) Well-being}

Human well-being was mentioned in relation to privacy (P13).
It was also described as a motivation for AI applications in controlling invasive species
that are human disease vectors~(P04).
The potential of harmful future applications of AI were mentioned
as issues for human well-being~(P20, P21).

Two participants mentioned concerns about applications of AI systems
that they regarded as having contested impacts on human and social well-being,
such as facial recognition~\cite{Buolamwini_2018,Raji_2020}
and dual-use applications~(P11, P12).
``[T]here is a spectrum of stuff where everybody has a comfort line [...].
It has nothing to do with privacy, fairness, and all those issues [mentioned in the principles].
It's a completely different topic''~(P11).

Uncertainty over the social impact of new technologies was also mentioned: 
``[t]he difficult thing really is to understand [...] how the technology really affects society.
It's not an easy thing to see''~(P12).
Concerns were also raised about data analyses and recommendations
that may have uneven benefits and reinforce disadvantage in society~(P01).

Environmental well-being was mentioned as a motivation for AI projects
that use environmental data~(P04, P09).
It was also identified as being significant for clients:
``environmental well-being was a big concern for them, 
not only because they are very environmentally concerned,
but also because that imposes requirements that some other customers [of the client] have''~(P07). 

AI projects may also contribute to environmental well-being by supporting environmental decision making: 
``the hope is that [...] in some cases we would be involved in helping people making decisions
when it matters the most [...] for a threatened species or ecosystem''~(P08).
However, this impact depends on the client following the recommendations of the AI system~(P08).

Social well-being was identified as an important issue for projects that involve Indigenous communities.
The information made available by AI systems helped Indigenous people
to collect information about dangerous areas in their land, 
and as a means of allowing them to maintain a connection to country
even if they physically remote from it~(P14).

\section{Discussion and Suggestions}
\label{sec:discussion}

In this section we discuss the observations and insights gained from the interviews
and place them in wider context,
with the aim of providing suggestions and caveats
for consideration when implementing the high-level AI ethics principles.
In the same vein as the preceding results section,
the discussion is organised into subsections,
covering each of the high-level principles.

\subsection{Privacy Protection and Security}
\label{sec:discussion_privacy}

This principle was largely interpreted as the protection of 
sensitive data and privacy across the lifespan of a system,
both during the development and deployment.
Sensitive data in this context may include 
data indicating racial or ethnic origin,
political and religious beliefs,
genetic and biometric data,
health-related data (including mental health),
sexual orientation,
financial data,
as well as criminal offences and convictions~\cite{GDPR_2016}.
Participants noted the importance of appropriate security methods to 
protect both the data and the AI system using the data.

Trust with collaborators and data providers is an important motivation for ensuring privacy and data protection.
Collaborators and data providers must have trust in the team and organisation
to not abuse the data and disclose their private information.

Privacy impact assessments and human research ethics committee reviews
are important tools for identifying privacy risks and consequences,
as well as resolving potential issues,
which may include recommendations on software engineering best practices.
However, it must be noted that the applicability of human research ethics committees is necessarily limited in scope,
as such committees typically only review research projects which directly involve humans.
It is possible that a given project may not perform research involving humans,
and still have privacy concerns.

Software engineering best practices may include the use of open algorithms
and open systems with a good security track record.
This may include proactive public disclosure and correction/patching of security vulnerabilities,
as opposed to the practice of \textit{security through obscurity}, 
where the algorithms and systems are kept secret/proprietary.
The efficacy of the latter practice has been questioned in the literature~\cite{Hoepman_2007,Mercuri_2003}.

Methods for protecting sensitive data in ML model training data may negatively affect the usefulness of the model. 
In general, the richer the data is, more accurate and/or reliable ML models can be built~\cite{Bishop_2006,Goodfellow_2016}.
On the other hand, privacy risks are likely to be reduced when less data is made available or used.
To fulfil privacy requirements, developers may need to aggregate the data
so that the risk of uniquely identifying individuals is reduced.
However, this has to be carefully weighed against the utility (accuracy and reliability) of the  ML models.

Privacy concerns differ from project to project,
where collaborators and clients may impose restrictions on data access by researchers. 
Access restrictions on sensitive data may prevent projects from proceeding,
and in some cases alternatives to sensitive data may be available.
Data requirements need to be associated with specific kinds of privacy measures
the customers/collaborators want to have in place.
For example, a customer/collaborator might be willing to share only coarse information
about the outcome of a process, excluding any finer details about the process 
(such as the steps of the process or data used within the process).

Much effort has been put into technologies that reduce the need to have complete access to training data.
For example, federated learning~\cite{TianLi_2020,QiangYang_2019} is a privacy preserving technique
and can be considered as an architectural design pattern
to deal with the trade-off between privacy and reliability,
as well as various requirements/restrictions placed by data providers.
Software architecture decisions are needed to ensure data privacy and security;
for example, which components read the data and which components share data outside the system.

Differential privacy~\cite{Dwork_2008,Friedman_2010} is a type of privacy-preserving method
that is intended to be robust against membership inference attacks
and attribute inference attacks which are specific for AI systems.
Membership inference can be used to deduce whether a certain individual's data
is used as part of the model training.
Attribute inference aims to infer missing parts of data for a given individual
by using partial data from the individual's record in conjunction with the output of an AI algorithm.
Differential privacy means the learned model should not depend too much on any individual's data.
The standard way to achieve the differential privacy property is to inject ``noise'' in the computation.
Differential privacy can be treated as an architectural design pattern for privacy;
examples include deliberate perturbation or random alteration of input data, model parameters, and/or output data.

More restricted access control should be applied to sensitive data (e.g., genomics data),
including only using a portion of data, never locating the data at the same space in the same time,
only sharing the insights of the data, encrypting the data while letting only the data owner own the decryption key.

Implementation of this principle may involve proper governance and management
for all data used and generated by AI systems~\cite{DISER_2020}.
Furthermore, appropriate security measures should be used for both the data and the AI system using the data.
This may include identification of potential security vulnerabilities within the underlying hardware and/or software;
open-source software may be preferable
due to not relying on the questionable \textit{security through obscurity} approach.
Security measures should also include appropriate mitigation measures for unintended applications of AI systems,
and/or potential abuse of AI systems.
Robustness to adversarial attacks should also be taken into account, 
where deliberately corrupted data is given to AI systems with the aim of confusing the systems
to produce unintended or adverse outcomes~\cite{Chakraborty_2021,Kumar_2020},
potentially reducing privacy and/or leaking sensitive data.

\subsection{Reliability and Safety}
\label{sec:discussion_reliability}

Participants interpreted this principle as meaning that the AI system
should produce accurate and reproducible recommendations.
The ability to produce reliable AI systems is often largely dependent
on the volume and quality of training data that is available~\cite{Bishop_2006,Goodfellow_2016}.
AI systems may be unreliable if they are used in a context
that differs from the one they are developed for.
AI systems are difficult to verify, and may require expert assessment
if they are used to replace an existing process or model.
Reliability and safety are critical for autonomous systems.

For situations where obtaining sufficient samples can be costly (in terms of time, effort, and/or financial cost),
\textit{active learning}~\cite{Settles_2012} may be beneficial and can be considered as a design pattern.
The overall goal in active learning is to find the most informative training samples
so that AI models can be successfully built from a relatively small number of training samples.
As an example of active learning, initial limited data (samples) is used to build an initial model that 
has the ability to indicate where (i.e.,~areas of sample space) it is most uncertain;
the locations with the highest uncertainty are then used to guide (either automatically or manually)
the selection of new samples to be obtained, which are then used to refine the model.
This iterative learning process is repeated until the model has sufficient fidelity
and/or reaching a cost threshold.

In designing of AI systems, it may be useful to explicitly consider 
unintended operation due to faulty implementation (including software bugs),
and erroneous/incomplete data (including training data and data used during deployment).
In both cases the system should be able to \textit{fail safely} by design (also known as \textit{fail-safe}).
This may include designing the system to refuse to provide an output (e.g., recommendation)
or explicitly marking the output as potentially unreliable.
The system should provide reasoning for the refusal and unreliable output,
such as incomplete input data, or a large mismatch between training data and given data.

The choice of the underlying algorithms 
should take into account at least three aspects and the trade-offs between them:
accuracy, robustness, and explainability.
The system should provide reliable outputs across a range of operating conditions (i.e.,~robustness)
expected during operation,
and not simply the best possible accuracy in a very narrow range (e.g., on a limited test dataset).
For example, maximising accuracy in well controlled settings may have detrimental impact
on accuracy within less controlled settings in actual deployment~\cite{Wong_2012},
including situations with missing, incomplete, or erroneous data.
Hence focusing only on accuracy in a narrow range of conditions increases risk of unreliable operation
in a wider range of conditions.
In application contexts where explainability is desired or required,
lower model complexity may be preferred,
at the possible cost of reduction in accuracy.

To achieve reliable functionality, the range of operating conditions may need to be explicitly limited,
possibly in tension with the \textit{Fairness} principle.
Training data may only be available for a set of operating conditions or specific subsets of people
(e.g., limited in age range, limited in ethnicity).
Operation beyond the training data may result in unreliable or erroneous outputs,
and in turn may result in loss of trust in the AI system.

For building trust in the reliability of an AI system, it may be useful to keep track of usage,
noting the suggestions made by the system and which suggestions are followed by the user
(cf. \textit{Contestability} principle).
This can be used to determine where the system works as designed,
and where the system needs to be improved.
This can also be used as an aid to improve the design of the system,
by taking into account new data and/or specifications that were initially not used.

To address performance degradation over time, 
regularly scheduled recalibration and/or retraining of AI models can be useful,
in order to take into account changes and/or improvements to the available data~\cite{Lewis_2022}.
Furthermore, automatic measurement of the nature and quality of input data
(e.g., missing values, intermittent availability, increase/decrease in rate of data)
can be used to automatically issue alerts that an AI system may not be performing as designed.

Implementation of this principle may involve adopting safety measures
that are proportionate to the magnitude of potential risks~\cite{DISER_2020}.
AI systems may need to be verified, validated and monitored on a continual basis
to ensure they maintain their operation for the intended purpose.
Continual risk management should be used to address any identified problems.
Responsibility for safety measures, monitoring, testing,
and risk management should be clearly and appropriately identified.

\subsection{Transparency and Explainability}
\label{sec:discussion_transparency}

While the principle is defined with the \textit{Transparency}
and \textit{Explainability} components having distinct meanings (see Table~\ref{tab:principles_summary}),
the two components were often observed to be interchangeable from the interviewees' perspective,
and interpreted often to mean only \textit{Explainability}.

Interpretability and explainability can treated as distinct requirements for AI systems.
An explanation of a model result is a description of how a model produces an outcome.
This explanation may need to be placed in terms that are understandable by the target audience;
there may be more than one target audience.
For example, deep neural networks can be completely explainable
in terms of relatively straightforward mathematics,
but the sheer amount of parameters is such networks makes them difficult to interpret~\cite{Arrieta_2020}.
An interpretable model should provide users with a description of what a stimulus,
such as a data point or model output, means in context.
The degree of required explainability should be a consideration in deciding
which algorithms should be used.

Interpretability is critical to building human trust in AI.
To improve interpretability, decision outcomes can be marked with a probability indicator
(e.g., 60\% accuracy estimate).
To improve trust, explainable interfaces can be designed to show graphical representation
of the decision making process and indicators of why a specific decision was made.
Explaining decisions with useful scenarios and the potential impact of various inputs
is helpful to improving trust in the system.
In other contexts, explainability may be a temporary requirement until users understand how the model works.
Once users understand how it works, transparency and explainability
may still be needed for verification and troubleshooting. 

Model outputs and explanations should be presented in ways
that relate to the users' background, culture and preferences.
The granularity of the explanation should also be considered when explaining decisions.
High-level explanations may be sufficient for some users and contexts,
while others may require more detailed explanations of the AI system's decision.

Thorough documentation is needed for explainability.
Reproducibility checklists that are already used by some venues for scientific publishing
overlap with the documentation requirements for explainability.
For example, one project team used a simple clustering algorithm
and recorded the decisions including the choice of clustering algorithm,
the choice of predictor variables,
how to represent categorical data numerically,
how to handle correlations between variables,
and how to normalise or scale data.
Sensitivity analysis is also useful to understand the sensitivities
and the uncertainty in decision making if there are resources to do so.
Accuracy (e.g., sensitivity and specificity) on training and testing data should also be documented.

A simplified model can be helpful for explaining how a complex model works
by giving an idea of the criteria affecting the complex model's decision making.
More specifically, 
improvements in explainability may be obtained
through the use of model approximation techniques
(including surrogate modelling and model emulation),
where simplified ML models are used to represent more complex models~\mbox{\cite{Asher_2015,Bolt_2023}}.
Here the aim is to produce a less complex model and/or more explainable model,
while preserving as much fidelity of the more complex model as possible.
One team developed algorithms/tools to simplify the complexity of solutions
by providing compact representation to find the most important components,
e.g., find a smaller decision tree that performs almost as well as the more complex one.
Furthermore,
Shapely values and Local Interpretable Model-Agnostic Explanations
are two model agnostic techniques
that can be used to aid explainability~\cite{nielsen_practical_2021}.

\subsection{Fairness}
\label{sec:discussion_fairness}

This principle addresses widespread concerns about the potential for AI systems 
to reflect (and potentially reinforce) existing inequalities within society. 
This covers both the accessibility of AI systems to users with diverse abilities 
and cultural backgrounds, and how decisions made based on the recommendations of 
these systems may disproportionately affect disadvantaged groups.
Decisions made  with the assistance of AI systems also need to comply
with anti-discrimination laws~\cite{DISER_2020}.

Concerns about fairness cover both biases embedded within the data used to train ML models,
and biases embedded in AI systems themselves~\cite{Dawson_2019}. 
These biases may be pre-existing individual and societal biases of those 
developing the system, technical biases resulting from how the system is implemented,
or emergent biases resulting from the context within which the system is used~\cite{friedman_bias_1996}.

In the interviews,
fairness was understood as decisions made on unbiased data, as non-discrimination,
and equal distribution of benefits and burdens of the output of AI systems.

There are multiple definitions and measures of fairness,
and choosing between them is both a technical problem and a question of values~\cite{Wong_2020}.
Algorithms cannot satisfy multiple definitions of fairness at the same time~\cite{Wong_2020}.
The appropriate definition and measure of fairness for an AI project should be discussed
between the project team, clients, and stakeholders at the start of the project.

Adjustments to ML models to promote fairness may occur in the input data for the model (pre-processing),
in the processing performed by the model,
or in the output of the model (post-processing)~\cite{nielsen_practical_2021}.
Software libraries that implement bias mitigation methods should be considered
if they are appropriate for the methods, tools, and goals of AI projects~\cite{Bellamy_2018}.

A record should be kept of the methods used for pre-processing (or cleaning) data
that is then used as inputs the ML model~\cite{Gebru_2021}.
This includes data used for training the model,
as well as data used during deployment of the model~\cite{Mitchell_2019}.
The methods include approaches for inferring missing values (or how missing values in data are handled),
which components (fields) of data are explicitly omitted,
which components of data are explicitly used,
and the specific approaches employed for mitigation of various biases
(e.g., placing weights on data during training and deployment).
See also the discussion on the \textit{Accountability} principle in Section~\ref{sec:discussion_accountability},
which proposes a related \textit{reproducibility checklist} concept.

If the data used for training and/or deployment of ML models comes from a third-party source, 
the data should not be used blindly.
The source should be queried about any known biases and potential biases within the data,
and how the data was collected.
Knowledge of the biases in the data can help with the design and application of bias mitigation approaches.

\subsection{Contestability}
\label{sec:discussion_contestability}

Contestability can be considered as critical for building trust in AI.
Those affected by either the output of an AI system
or by decisions made based on the output of these systems 
should have the opportunity to question these outcomes and correct them if they are in error.
In particular, vulnerable groups should be considered during the development and use of AI systems. 

Contestability may be implemented by allowing users to override the recommendations and decisions of AI systems.
It may be difficult to revise the initial assumptions and design decisions in the system
if they are contested by those affected.
The explanation of the AI system's decision process may also be too complex for users to understand
(cf. \textit{Transparency and Explainability} principle),
or the relationships in the data may not be well understood.
Contestability was also understood as interpretability.

The explainability of an AI system is important for specific aspects of how it produced the output to be contested.
In addition to satisfying explainability requirements to those affected by the AI system
(e.g., descriptions of the system and underlying algorithms),
information on how to contest the system's output should also be provided. 
The means of obtaining this information and how to contest this output
may be included with the description of the output provided to those affected by it.

The ability to opt-in or opt-out of AI systems should be provided to users if
possible. More specifically, an explicit option may be provided to the user to 
disagree with decisions given by an AI system. 

Where the output of the AI system is to augment an expert's decision making,
it may be necessary to confirm whether regulatory approval is necessary for such 
decision support. This may be the case if the system is assisting decisions that 
are diagnostic or therapeutic in nature.

\subsection{Accountability}
\label{sec:discussion_accountability}

The participants interpreted this principle as accountability for the methods and data used by an AI system,
as well as the outputs of the system.

Accountability relates to transparency and explainability;
giving an explanation for an AI system was seen as a requirement for being accountable for it.
Evaluating the system through test data is important for accountability.
There is no consensus on who is accountable for the outcomes of decisions made based on the system's recommendations.
Accountability is also associated with reliability and safety,
and is seen as important for the acceptance of autonomous systems.
For indigenous collaborators and co-creators,
accountability extends beyond the outputs of the system itself
to the benefits of the larger process that the AI system is embedded within.

A possible complicating factor in accountability is that some AI systems may continue to learn post-deployment,
using, for example, data provided by users.
If there are issues with the new data, the accountability boundary is blurred between
the original developers of such systems and the users of the systems.

Accountability is expected to be role-level and associated with traceability and reproducibility.
AI systems may provide end users a method to track their historical use of the system
(e.g., to discover faults through an immutable log) and allow them to provide feedback.
Handover points in the development process may be identified to ensure traceability and accountability,
as handover points are where traceability may be lost. 

If a method developed by design works for a particular client for a specific dataset,
it is challenging to maintain reproducibility. In current practice,
project leaders have to describe why they want to use data,
how they are going to use it,
what they are going to do with the data,
what are the risks,
what is the consent,
who is going to have access to that data,
for how long you are going to keep it,
and what happens to the data at the end of the project.
This current practice may be augmented by introducing a \textit{reproducibility checklist}
that provides the information necessary to reproduce the output of an AI system.
This checklist may include:

\begin{enumerate}[{$\bullet$}]
\itemsep=0.2ex
\item how the data was pre-processed
\item how the data was scaled or normalised
\item the approaches used to mitigate biases (see Section~\ref{sec:discussion_fairness})
\item how hyperparameters were configured in the model
\item the algorithms chosen for use in the system
\item the choice of prediction variables
\item the numerical representation of categorical data
\item how correlation between variables was handled
\end{enumerate}

Domain knowledge needs to be reflected within the AI system.
There may be  various versions of domain knowledge frameworks
that may be useful to employ in AI systems that operate in specialised domains. 

Collaboration and co-creation of AI projects with groups
who are end-users or significantly affected by the system's output
is a powerful means of making these systems accountable to them.
Techniques from participatory design, co-design (collaborative design)
and citizen-led projects may be useful to consider for AI projects
with significant community impacts~\cite{simonsen_design_2013, southern_2014, whittle_community_2020}.

\subsection{Human-Centred Values}

Human values within projects are rarely stated explicitly,
and may not be apparent within projects where the AI system does not interact with human users.
Principles such as \textit{Fairness} and \textit{Privacy}
can be useful for ensuring that AI systems are aligned with human values.
AI projects co-created with Indigenous people should reflect Indigenous worldviews and knowledge practices.

Human values have been well studied within software engineering~\cite{Whittle_2019,Hussain_2020,Whittle_2021},
where Schwartz's well-evidenced set of universal values are used~\cite{Schwartz_2012a,Schwartz_2012b}. 
The resulting approaches can be adapted to address human values and concerns in AI systems.
Human-centred values may also be partially addressed by fostering diversity within development teams,
as a wide range of perspectives can identify potential concerns about how the system may affect people's rights.

Indigenous cultural and intellectual property rights knowledge sharing protocols and practices
need to be respected and protected.
As such, Indigenous people should co-design AI projects to ensure the incorporation
of Indigenous knowledge systems and governance systems.
The CARE principles (Collective Benefit, Authority to Control, Responsibility and Ethics)
for Indigenous data governance~\cite{CARE_2019},
and the \textit{Our Knowledge Our Way} best practice guidelines~\cite{woodward_our_2020}
can enable the combination of Indigenous knowledge, science and AI system design to be woven together,
in order to empower Indigenous people and decision-making.
Examples of how to translate CARE principles into AI system design are given in~\cite{Robinson_2021}.
In turn, this may help with the implementation of other principles,
such as \textit{Privacy Protection and Security}, and \textit{Accountability}.

It may be necessary to perform due diligence into potential reputational risks for an organisation
by working with particular business partners in controversial applications of AI systems,
such as dual-use and defence applications.

\subsection{Human, Social and Environmental (HSE) Well-being}
\label{sec:discussion_wellbeing}

Existing and potential future applications of AI were mentioned as a risk for human well-being.
Privacy was mentioned as an aspect of human well-being.
Uncertainty about the social impact of new technologies was also mentioned.
Environmental well-being was seen as a motivation for projects that use environmental data,
and was seen as important to clients.
Social well-being was an important aspect of projects that involve Indigenous people.
The effects of deploying an AI system may be recorded
to better understand the benefits gained and burdens imposed by introducing the system. 

Respect for culture is important in AI projects,
where there may be race- or group-specific aspects to data, model outputs or decisions.
Cultural concerns may be incorporated into both functional and data requirements for AI projects.
For Indigenous projects, Indigenous people need to be involved in environmental decision-making.
For example, critical environmental decisions can take into account Indigenous knowledge.

\vspace{-1ex}
\section{Support Mechanisms}
\label{sec:recommendations}

Based on the insights primarily synthesised from our interview study,
below we provide two sets of suggestions
that may help with the implementation of the high-level ethical principles into practice.
The first set covers organisational support,
while the second set suggests the use of design notes as a companion to high-level ethics principles.

\vspace{-1ex}
\subsection{Organisational Support}

For organisations that \textit{want to be ethical} with AI,
devising an implementation process of AI ethics principles 
is a step towards building capacity and capability for \textit{being ethical} with AI.
The potential benefits may include 
gained trust leading to competitive advantage, 
retention of highly-skilled staff, 
avoidance of reputational damage, 
as well as responsiveness and readiness for regulatory requirements.

Adoption of responsible AI as a fundamental aspect of an organisation's digital strategy
may be beneficial for long term development and growth.
Within organisations or departments devoted to scientific research,
a digital strategy can include investment in the responsible development and use of AI
to accelerate scientific discovery through the use of current and future digital technologies
(e.g., digital assistants, automation and robotics).

It may be useful to provide organisation-wide training and processes
to increase awareness and understanding of high-level AI principles
followed by organisation-wide adoption and implementation of such principles.
This can be accomplished through a set of seminars,
and/or recommended online learning modules.
The chosen principles can be the generally accepted AI ethics principles~\cite{Jobin_2019},
or a specific set of principles such as the Australian AI ethics principles~\cite{DISER_2020},
as summarised in Table~\ref{tab:principles_summary}.

Several interviewees noted that the high-level principles do not provide guidance
as to where AI should and should not be used
(e.g., in projects with dual-use applications and/or non-civilian focus).
As such, it may be useful to define a specific organisational policy on the permissible uses of AI systems.

Analysis of existing software, systems, and/or projects that use AI technologies
may yield areas where consideration and/or implementation of the AI ethics principles
can be advantageous and/or prudent.
In addition to looking at internally produced software and systems,
liaison with external software producers/providers
(which includes closed-source/proprietary software as well as open-source software)
to ensure software used within an organisation adheres to AI ethics principles.
This may have a useful and far-reaching side effect in that the high-level AI ethics principles
operationalised in externally produced software can positively affect the functioning of many other organisations (on a global scale) which also use the same software.

It may be beneficial to start each project with a Data Planner (RDP)~\cite{ARDC_2022},
where the need and/or use of AI technologies can be identified.
This allows the identification of applicable high-level AI ethics principles within the context of each project,
which in turn spurs the formulation of plans for addressing/implementing each applicable principle.
The RDP may be helpful in providing links to useful material that raises awareness of the principles;
it may also be helpful for building a vignette to facilitate addressing applicable principles.
As a useful side effect, use of the RDP may lead to an increase in the organisational knowledge
of the AI ethics principles,
as well as an increase in the pool of methods and approaches for operationalising each principle.

Related to RDP,
the application of \textit{dataset datasheets}~\cite{Gebru_2021}
is encouraged whenever a dataset is created and/or an existing dataset used
for AI models, projects and/or products.
Such datasheets cover characteristics of given datasets,
such as their motivation, composition, collection/acquisition methods,
applied cleaning/pre-processing, added labels, recommended uses, etc.
Given that datasets are an integral part of the development and use of AI systems,
the use of dataset datasheets can increase accountability and transparency.
In the same vein, use of \textit{model cards}~\cite{Mitchell_2019}
is also encouraged in order to describe the characteristics of AI models,
such as their type, version, intended use, performance (error rates and associated metrics),
as well as the names and versions of associated datasets used for training and evaluation of the models.

\vspace{-1ex}
\subsection{Design Notes for Developers}

Current high-level principles, such as the Australian AI ethics principles~\cite{DISER_2020},
appear more focused on end users and people affected by AI technologies,
rather than developers of AI systems.
This results in a mismatch between the underlying intent of the principles
and how the principles are to be implemented.
To address this mismatch, the high-level principles can be extended
with a {mapping between the present principles and a set of \textit{design notes} aimed at developers},
such as suggestions and specific attributes
to keep in mind while designing and implementing AI systems.

Examples of suggestions that can be part of the design notes are given in the discussion
in Section~\ref{sec:discussion}.
We note that the listed approaches reflect the content of the interviews 
and the surrounding discussions for each principle,
and are hence not exhaustive.
As~such, including recommendations and suggestions
from relevant literature~\cite{Mantymaki_2022,Raji_2020b,Seppala_2021,Shneiderman_2020,Shneiderman_2021}
may yield a richer picture.

The proposed design notes may also include reusable design methods,
where the design and implementation of system components,
as well as their integration, 
follows known process/design patterns that are aligned with high-level AI ethics principles~\cite{Lu_2022,Lu_2024}.

Within the Australian AI ethics principles~\cite{DISER_2020},
the \textit{Privacy Protection and Security} principle covers seemingly closely related topics.
However, in practice the two notions of security and privacy are distinct,
and can be treated as two separate principles in the proposed design notes
in order to better match developers' understanding and practice.
Security in this context typically means security at the software and deployment level,
where best practices in software engineering and deployment can be used.
Privacy protection in the context of AI systems involves the careful treatment of sensitive data,
where approaches specific to AI need to be used.
This includes approaches such as
federated machine learning, 
deidentification,
and differential privacy.
Using these approaches may affect the utility (e.g., accuracy and reliability) of AI systems,
and hence must be carefully considered and weighed against the degree of required privacy protection.

A related compounding problem exists in the \textit{Transparency and Explainability} principle,
where the two components have distinct meanings.
Yet from the developers' perspective, the two components were often observed to be interchangeable
and interpreted often to mean \textit{Explainability},
resulting in the \textit{Transparency} component being de-emphasised.
In order to address this downside, the \textit{Transparency} component
can be separated into a separate stand-alone principle within the proposed design notes.
While it may be tempting to join \textit{Transparency} with the related \textit{Contestability} principle,
from developers' perspective these two principles are distinct tasks and hence require separate solutions.
In a similar vein, the \textit{Reliability and Safety} principle covers two related yet distinct issues,
which may be better treated separately.

\section{Conclusion}
\label{sec:conclusion}

In this interview study, the unique requirements, constraints and objectives of 
the diverse range of projects that participants have been involved in provide 
valuable insights into the complexities of designing and developing responsible, 
or ethical, AI systems.
Using high-level AI principles proposed by the Australian Government
provided a structure for the analysis and discussions,
which in turn yielded insight into how the ethics principles
were interpreted in the context of the professional experience of the participants.

A notable challenge that emerged was balancing the tensions and \mbox{trade-offs}
that were inherent to using the principles%
\footnote
  {
  In follow-up work we have examined in more detail various trade-offs
  and frameworks for resolving them~\cite{Sanderson_2023,Sanderson_2024}.
  }%
.
Tensions were highlighted between the practical approaches used
for implementing privacy and security, transparency and explainability, as well as accuracy.
For example, techniques may be used to develop an AI system that provides
highly accurate outputs with the caveat that those outputs are not explicitly explainable.
Similarly, the development of a highly explainable AI system may be vulnerable to privacy risks.
In these cases, a choice can be made to prioritise one set of values over another~\cite{Whittlestone_2019}.

Weighing the risks and benefits of such decisions
requires the implementation of supportive mechanisms
and should not fall to the designers and developers of AI systems alone~\cite{Seppala_2021}.
The ability to assess risk was also presented as critical across the interviews
as a way to gauge the degree of oversight, accountability, reliability and explainability
required for an AI system. 

We note that the existence of the principles alone does not guarantee ethical AI~\cite{Mittelstadt_2019}.
Further infrastructure, beyond the organisational level,
may be required  to build the capability and capacity of AI developers and designers
to create responsible AI~\cite{Seppala_2021}.
High-level support and governance of AI development to support 
the practice of implementing principles is required.
Such support could be in the form of professional codes and regulatory frameworks,
as well as legal and professional accountability mechanisms
to uphold professional standards and provide users
with redress for negligent behaviour~\cite{Mittelstadt_2019}.

\section*{Acknowledgements}

We would like to thank our colleagues (Andreas Duenser, Liming Zhu, Ross Darnell, Dan Gladish)
for reviewing drafts of this work.
We would also like to thank the anonymous reviewers for comments and suggestions
that led to the improvements of this work.

\vspace{1ex}


\bibliographystyle{ieee_mod}
\bibliography{references}

\begin{thebibliography}{10}\itemsep=0.40ex

\bibitem{Asher_2015}
M.~J. Asher, B.~F.~W. Croke, A.~J. Jakeman, and L.~J.~M. Peeters.
\newblock A review of surrogate models and their application to groundwater
  modeling.
\newblock {\em Water Resources Research}, 51(8):5957--5973, 2015.

\bibitem{DISER_2020}
{Australian~Government~Dept.~Industry,~Science~and~Resources}.
\newblock {Australia's Artificial Intelligence Ethics Framework}, 2019.
\newblock Online.~{URL:}
  https://industry.gov.au/publications/australias-artificial-intelligence-ethics-framework
  (accessed: 12 Apr 2023).

\bibitem{ARDC_2022}
{Australian Research Data Commons}.
\newblock Data management plans, 2022.
\newblock Online. {URL:}
  {https://ardc.edu.au/resources/aboutdata/data-management-plans} (accessed: 21
  Sep 2022).

\bibitem{Arrieta_2020}
A.~Barredo~Arrieta, N.~Díaz-Rodríguez, J.~Del~Ser, A.~Bennetot, S.~Tabik,
  A.~Barbado, S.~Garcia, S.~Gil-Lopez, D.~Molina, R.~Benjamins, R.~Chatila, and
  F.~Herrera.
\newblock Explainable artificial intelligence ({XAI}): Concepts, taxonomies,
  opportunities and challenges toward responsible {AI}.
\newblock {\em Information Fusion}, 58:82--115, 2020.

\bibitem{Bellamy_2018}
R.~K.~E. Bellamy, K.~Dey, M.~Hind, S.~C. Hoffman, S.~Houde, et~al.
\newblock {AI Fairness} 360: An extensible toolkit for detecting,
  understanding, and mitigating unwanted algorithmic bias.
\newblock \textit{arXiv:1810.01943}, 2018.

\bibitem{Bishop_2006}
C.~M. Bishop.
\newblock {\em Pattern Recognition and Machine Learning}.
\newblock Springer, 2006.

\bibitem{Bolt_2023}
A.~Bolt, C.~Sanderson, J.~J. Dabrowski, C.~Huston, and P.~\mbox{Kuhnert}.
\newblock A~neural emulator for uncertainty estimation of fire propagation.
\newblock {\em \mbox{Procedia} Computer Science}, 222:367--376, 2023.

\bibitem{Braun_2006}
V.~Braun and V.~Clarke.
\newblock Using thematic analysis in psychology.
\newblock {\em Qualitative Research in Psychology}, 3(2):77--101, 2006.

\bibitem{Braun_2019}
V.~Braun, V.~Clarke, N.~Hayfield, and G.~Terry.
\newblock {Thematic Analysis}.
\newblock In {\em {Handbook of Research Methods in Health Social Sciences}},
  pages 843--860. Springer, 2019.

\bibitem{Buolamwini_2018}
J.~Buolamwini and T.~Gebru.
\newblock Gender shades: Intersectional accuracy disparities in commercial
  gender classification.
\newblock In {\em Conference on Fairness, Accountability and Transparency},
  pages 77--91, 2018.

\bibitem{Corinne_2018}
C.~Cath.
\newblock Governing artificial intelligence: ethical, legal and technical
  opportunities and challenges.
\newblock {\em Philosophical Transactions of the Royal Society A: Mathematical,
  Physical and Engineering Sciences}, 376(2133):20180080, 2018.

\bibitem{Chakraborty_2021}
A.~Chakraborty, M.~Alam, V.~Dey, A.~Chattopadhyay, and D.~Mukhopadhyay.
\newblock A survey on adversarial attacks and defences.
\newblock {\em {CAAI} Transactions on Intelligence Technology}, 6(1):25--45,
  2021.

\bibitem{Christoforaki_2022}
M.~Christoforaki and O.~Beyan.
\newblock {AI} ethics -- a bird's eye view.
\newblock {\em Applied Sciences}, 12(9):4130, 2022.

\bibitem{Coeckelbergh_2020}
M.~Coeckelbergh.
\newblock {\em AI Ethics}.
\newblock The MIT Press, Cambridge, Massachusetts, 2020.

\bibitem{Colvin_2016}
R.~Colvin, G.~B. Witt, and J.~Lacey.
\newblock Approaches to identifying stakeholders in environmental management:
  Insights from practitioners to go beyond the `usual suspects'.
\newblock {\em Land Use Policy}, 52:266--276, 2016.

\bibitem{dAquin_2018}
M.~d'Aquin, P.~Troullinou, N.~E. O'Connor, A.~Cullen, G.~Faller, and L.~Holden.
\newblock Towards an `ethics by design' methodology for {AI} research projects.
\newblock In {\em AAAI/ACM Conference on AI, Ethics, and Society}, pages
  54--59, 2018.

\bibitem{Dawson_2019}
D.~Dawson, E.~Schleiger, J.~Horton, J.~McLaughlin, C.~Robinson, G.~Quezada,
  J.~Scowcroft, and S.~Hajkowicz.
\newblock Artificial intelligence: Australia's ethics framework.
\newblock {Report EP191846}, {CSIRO}, 2019.

\bibitem{Dignum_2019}
V.~Dignum.
\newblock {\em Responsible Artificial Intelligence}.
\newblock Springer International Publishing, 2019.

\bibitem{Dwork_2008}
C.~Dwork.
\newblock Differential privacy: A survey of results.
\newblock {\em Lecture Notes in Computer Science (LNCS)}, 4978:1--19, 2008.

\bibitem{Eubanks_2018}
V.~Eubanks.
\newblock {\em Automating Inequality: How High-Tech Tools Profile, Police, and
  Punish the Poor}.
\newblock Picador, New York, NY, 2018.

\bibitem{EC_2021}
{European Commission}.
\newblock {Proposal for a Regulation of the European Parliament and of the
  Council Laying Down Harmonised Rules on Artificial Intelligence (Artificial
  Intelligence Act) and Amending Certain Union Legislative Acts}, 2021.
\newblock {Document~52021PC0206~(1)}.

\bibitem{GDPR_2016}
{European Union}.
\newblock {Regulation (EU) 2016/679 of the European Parliament and of the
  Council of 27 April 2016 on the protection of natural persons with regard to
  the processing of personal data and on the free movement of such data, and
  repealing Directive 95/46/EC (General Data Protection Regulation)}.
\newblock {\em Official Journal of the European Union L~119}, 4~May~2016.

\bibitem{Farthing_2021}
S.~Farthing, J.~Howell, K.~Lecchi, Z.~Paleologos, P.~Saintilan, and E.~Santow.
\newblock {Human Rights and Technology}.
\newblock {Report}, {Australian Human Rights Commission}, 2021.
\newblock Online. {URL:}
  {https://humanrights.gov.au/our-work/rights-and-freedoms/publications/human-rights-and-technology-final-report-2021}
  (accessed: 14~Dec~2021).

\bibitem{Ferrario_2017}
M.~A. Ferrario, W.~Simm, J.~Whittle, C.~Frauenberger, G.~Fitzpatrick, and
  P.~Purgathofer.
\newblock Values in computing.
\newblock In {\em CHI Conference Extended Abstracts on Human Factors in
  Computing Systems}, pages 660--667, 2017.

\bibitem{Fjeld_2020}
J.~Fjeld, N.~Achten, H.~Hilligoss, A.~C. Nagy, and M.~Srikumar.
\newblock Principled artificial intelligence: Mapping consensus in ethical and
  rights-based approaches to principles for {AI}.
\newblock {Berkman Klein Center for Internet \& Society at Harvard University},
  Research Publication No.~2020-1, 2020.

\bibitem{Friedman_2010}
A.~Friedman and A.~Schuster.
\newblock Data mining with differential privacy.
\newblock In {\em International Conference on Knowledge Discovery and Data
  Mining}, pages 493--502, 2010.

\bibitem{friedman_bias_1996}
B.~Friedman and H.~Nissenbaum.
\newblock Bias in computer systems.
\newblock {\em ACM Transactions on Information Systems}, 14(3):330--347, 1996.

\bibitem{Gebru_2021}
T.~Gebru, J.~Morgenstern, B.~Vecchione, J.~W. Vaughan, H.~Wallach,
  H.~{Daum\'{e}~III}, and K.~Crawford.
\newblock Datasheets for datasets.
\newblock {\em Communications of the {ACM}}, 64(12):86--92, 2021.

\bibitem{Goodfellow_2016}
I.~Goodfellow, Y.~Bengio, and A.~Courville.
\newblock {\em Deep Learning}.
\newblock MIT Press, 2016.

\bibitem{Hajkowicz_2019}
S.~Hajkowicz, S.~Karimi, T.~Wark, C.~Chen, M.~Evans, N.~Rens, D.~Dawson,
  A.~Charlton, T.~Brennan, C.~Moffatt, S.~Srikumar, and K.~Tong.
\newblock Artificial intelligence: Solving problems, growing the economy and
  improving our quality of life.
\newblock {Report EP191848}, {CSIRO}, 2019.

\bibitem{Hall_2015}
N.~Hall, J.~Lacey, S.~Carr-Cornish, and A.-M. Dowd.
\newblock Social licence to operate: understanding how a concept has been
  translated into practice in energy industries.
\newblock {\em Journal of Cleaner Production}, 86:301--310, 2015.

\bibitem{Hoepman_2007}
J.-H. Hoepman and B.~Jacobs.
\newblock Increased security through open source.
\newblock {\em Communications of the {ACM}}, 50(1):79--83, 2007.

\bibitem{Hussain_2020}
W.~Hussain, H.~Perera, J.~Whittle, A.~Nurwidyantoro, R.~Hoda, R.~A. Shams, and
  G.~Oliver.
\newblock Human values in software engineering: Contrasting case studies of
  practice.
\newblock {\em IEEE Transactions on Software Engineering}, 48(5):1818--1833,
  2022.

\bibitem{Jobin_2019}
A.~Jobin, M.~Ienca, and E.~Vayena.
\newblock The global landscape of {AI} ethics guidelines.
\newblock {\em Nature Machine Intelligence}, 1(9):389--399, 2019.

\bibitem{Koopman_2020}
B.~Koopman, D.~Bradford, and D.~Hansen.
\newblock Exemplars of artificial intelligence and machine learning in
  healthcare.
\newblock {Report EP203543}, {CSIRO}, 2020.

\bibitem{Lewis_2022}
G.~A. Lewis, S.~Echeverría, L.~Pons, and J.~Chrabaszcz.
\newblock Augur: A step towards realistic drift detection in production {ML}
  systems.
\newblock In {\em IEEE/ACM International Workshop on Software Engineering for
  Responsible Artificial Intelligence (SE4RAI)}, pages 37--144, 2022.

\bibitem{TianLi_2020}
T.~Li, A.~K. Sahu, A.~Talwalkar, and V.~Smith.
\newblock Federated learning: Challenges, methods, and future directions.
\newblock {\em IEEE Signal Processing Magazine}, 37(3):50--60, 2020.

\bibitem{Liu_2021}
N.~Liu, P.~Shapira, and X.~Yue.
\newblock Tracking developments in artificial intelligence research:
  constructing and applying a new search strategy.
\newblock {\em Scientometrics}, 126(4):3153--3192, 2021.

\bibitem{Lu_2022}
Q.~Lu, L.~Zhu, X.~Xu, J.~Whittle, D.~Douglas, and C.~\mbox{Sanderson}.
\newblock Software engineering for responsible {AI}: An empirical study and
  operationalised patterns.
\newblock In {\em IEEE/ACM International Conference on Software Engineering:
  Software Engineering in Practice (ICSE-SEIP)}, 2022.

\bibitem{Lu_2024}
Q.~Lu, L.~Zhu, X.~Xu, J.~Whittle, D.~Zowghi, and A.~Jacquet.
\newblock \mbox{Responsible} {AI} pattern catalogue: A collection of best
  practices for {AI} governance and engineering.
\newblock {\em ACM Computing Surveys}, 56(7):1--35, 2024.

\bibitem{Mantymaki_2022}
M.~M\"{a}ntym\"{a}ki, M.~Minkkinen, T.~Birkstedt, and M.~Viljanen.
\newblock Defining organizational {AI} governance.
\newblock {\em AI and Ethics}, 2(4):603--609, 2022.

\bibitem{Mercuri_2003}
R.~T. Mercuri and P.~G. Neumann.
\newblock Security by obscurity.
\newblock {\em Communications of the {ACM}}, 46(11):160, 2003.

\bibitem{Mikalef_2022}
P.~Mikalef, K.~Conboy, J.~E. Lundstr\"{o}m, and A.~Popovi\^{c}.
\newblock Thinking responsibly about responsible {AI} and `the dark side' of
  {AI}.
\newblock {\em {European Journal of Information Systems}}, 31(3):257--268,
  2022.

\bibitem{Mitchell_2019}
M.~Mitchell, S.~Wu, A.~Zaldivar, P.~Barnes, L.~Vasserman, B.~Hutchinson,
  E.~Spitzer, I.~D. Raji, and T.~Gebru.
\newblock Model cards for model reporting.
\newblock In {\em Conference on Fairness, Accountability, and Transparency},
  pages 220--229, 2019.

\bibitem{Mittelstadt_2019}
B.~Mittelstadt.
\newblock Principles alone cannot guarantee ethical {AI}.
\newblock {\em Nature Machine Intelligence}, 1(11):501--507, 2019.

\bibitem{Morley_2021b}
J.~Morley, A.~Elhalal, F.~Garcia, L.~Kinsey, J.~M\"{o}kander, and L.~Floridi.
\newblock Ethics as a service: A pragmatic operationalisation of {AI} ethics.
\newblock {\em Minds \& Machines}, 31(2):239--256, 2021.

\bibitem{Morley_2020}
J.~Morley, L.~Floridi, L.~Kinsey, and A.~Elhalal.
\newblock From what to how: An initial review of publicly available {AI} ethics
  tools, methods and research to translate principles into practices.
\newblock {\em Science and Engineering Ethics}, 26(4):2141--2168, 2020.

\bibitem{Morley_2021a}
J.~Morley, L.~Kinsey, A.~Elhalal, F.~Garcia, M.~Ziosi, and L.~Floridi.
\newblock Operationalising {AI} ethics: barriers, enablers and next steps.
\newblock {\em AI \& Society}, 2021.

\bibitem{NMHRC_2018}
{National Health and Medical Research Council (NHMRC)}.
\newblock {National Statement on Ethical Conduct in Human Research - Updated},
  2018.
\newblock Online. {URL:}
  {https://www.nhmrc.gov.au/about-us/publications/national-statement-ethical-conduct-human-research-2007-updated-2018}.

\bibitem{nielsen_practical_2021}
A.~Nielsen.
\newblock {\em Practical {Fairness}}.
\newblock O'Reilly Media, Sebastopol, CA, 2021.

\bibitem{ONeil_2016}
C.~O'Neil.
\newblock {\em Weapons of Math Destruction: How Big Data Increases Inequality
  and Threatens Democracy}.
\newblock Allen Lane, London, 2016.

\bibitem{Raji_2020}
I.~D. Raji, T.~Gebru, M.~Mitchell, J.~Buolamwini, J.~Lee, and E.~Denton.
\newblock Saving face: Investigating the ethical concerns of facial recognition
  auditing.
\newblock In {\em AAAI/ACM Conference on AI, Ethics, and Society}, pages
  145--151, 2020.

\bibitem{Raji_2020b}
I.~D. Raji, A.~Smart, R.~N. White, M.~Mitchell, T.~Gebru, B.~Hutchinson,
  J.~Smith-Loud, D.~Theron, and P.~Barnes.
\newblock Closing the {AI} accountability gap: Defining an end-to-end framework
  for internal algorithmic auditing.
\newblock In {\em Conference on Fairness, Accountability, and Transparency},
  2020.

\bibitem{CARE_2019}
{Research Data Alliance International Indigenous Data Sovereignty Interest
  Group}.
\newblock {CARE} {Principles} of {Indigenous} {Data} {Governance}.
\newblock {\em Global Indigenous Data Alliance}, 2019.
\newblock Online. {URL:} {https://www.gida-global.org/care} (accessed: 07 Sep
  2022).

\bibitem{Robinson_2021}
C.~J. Robinson, T.~Kong, R.~Coates, I.~Watson, C.~Stokes, P.~Pert,
  A.~McConnell, and C.~Chen.
\newblock Caring for indigenous data to evaluate the benefits of indigenous
  environmental programs.
\newblock {\em Environmental Management}, 68:160--169, 2021.

\bibitem{Ryan_2021}
M.~Ryan, J.~Antoniou, L.~Brooks, T.~Jiya, K.~Macnish, and B.~Stahl.
\newblock Research and practice of {AI} ethics: A case study approach
  juxtaposing academic discourse with organisational reality.
\newblock {\em Science and Engineering Ethics}, 27(2), 2021.

\bibitem{Sanderson_2023}
C.~Sanderson, D.~Douglas, and Q.~Lu.
\newblock Implementing responsible {AI}: Tensions and trade-offs between ethics
  aspects.
\newblock In {\em International Joint Conference on Neural Networks (IJCNN)},
  2023.

\bibitem{Sanderson_2024}
C.~Sanderson, E.~Schleiger, D.~Douglas, P.~Kuhnert, and Q.~Lu.
\newblock \mbox{Resolving} ethics trade-offs in implementing responsible {AI}.
\newblock In {\em IEEE Conference on Artificial Intelligence (CAI)}, pages
  1208--1213, 2024.

\bibitem{Schwartz_2012a}
S.~H. Schwartz.
\newblock {An Overview of the Schwartz Theory of Basic Values}.
\newblock {\em Online Readings in Psychology and Culture}, 2(1), 2012.

\bibitem{Schwartz_2012b}
S.~H. Schwartz, J.~Cieciuch, M.~Vecchione, E.~Davidov, R.~Fischer,
  C.~Beierlein, A.~Ramos, M.~Verkasalo, J.-E. L\"{o}nnqvist, K.~Demirutku,
  O.~Dirilen-Gumus, and M.~Konty.
\newblock Refining the theory of basic individual values.
\newblock {\em Journal of Personality and Social Psychology}, 103(4):663--688,
  2012.

\bibitem{Seppala_2021b}
A.~Sepp\"{a}l\"{a}.
\newblock Implementing ethical {AI}: From principles to {AI} governance.
\newblock Master's thesis, Turku School of Economics, University of Turku,
  2021.

\bibitem{Seppala_2021}
A.~Sepp\"{a}l\"{a}, T.~Birkstedt, and M.~M\"{a}ntym\"{a}ki.
\newblock From ethical {AI} principles to governed {AI}.
\newblock In {\em International Conference on Information Systems}, 2021.

\bibitem{Settles_2012}
B.~Settles.
\newblock {\em Active Learning}.
\newblock Morgan \& Claypool, 2012.

\bibitem{Shneiderman_2020}
B.~Shneiderman.
\newblock Bridging the gap between ethics and practice: Guidelines for
  reliable, safe, and trustworthy human-centered {AI} systems.
\newblock {\em ACM Transactions on Interactive Intelligent Systems},
  10(4):1--31, 2020.

\bibitem{Shneiderman_2021}
B.~Shneiderman.
\newblock Responsible {AI}: Bridging from ethics to practice.
\newblock {\em Communications of the ACM}, 64(8):32--35, 2021.

\bibitem{simonsen_design_2013}
J.~Simonsen and T.~Robertson, editors.
\newblock {\em Routledge International Handbook of Participatory Design}.
\newblock Routledge, London and New York, 2013.

\bibitem{Kumar_2020}
R.~S. Siva~Kumar, M.~Nystr\"{o}m, J.~Lambert, A.~Marshall, M.~Goertzel,
  A.~Comissoneru, M.~Swann, and S.~Xia.
\newblock Adversarial machine learning -- industry perspectives.
\newblock In {\em {IEEE} Security and Privacy Workshops}, 2020.

\bibitem{southern_2014}
J.~Southern, R.~Ellis, M.~A. Ferrario, R.~McNally, R.~Dillon, W.~Simm, and
  J.~Whittle.
\newblock Imaginative labour and relationships of care: Co-designing prototypes
  with vulnerable communities.
\newblock {\em Technological Forecasting and Social Change}, 84:131--142, 2014.

\bibitem{Stahl_2021}
B.~C. Stahl, J.~Antoniou, M.~Ryan, K.~Macnish, and T.~Jiya.
\newblock Organisational responses to the ethical issues of artificial
  intelligence.
\newblock {\em AI \& Society}, 37(1):23--37, 2022.

\bibitem{Whittle_2019}
J.~Whittle.
\newblock Is your software valueless?
\newblock {\em IEEE Software}, 36(3):112--115, 2019.

\bibitem{whittle_community_2020}
J.~Whittle, M.~A. Ferrario, and W.~Simm.
\newblock {Community-University Research: A Warts and All Account}.
\newblock In {\em {Into the Wild: Beyond the Design Research Lab}}, pages
  115--147. Springer, 2020.

\bibitem{Whittle_2021}
J.~Whittle, M.~A. Ferrario, W.~Simm, and W.~Hussain.
\newblock A case for human values in software engineering.
\newblock {\em IEEE Software}, 38(1):106--113, 2021.

\bibitem{Whittlestone_2019}
J.~Whittlestone, R.~Nyrup, A.~Alexandrova, and S.~Cave.
\newblock The role and limits of principles in {AI} ethics: Towards a focus on
  tensions.
\newblock In {\em AAAI/ACM Conference on AI, Ethics, and Society}, 2019.

\bibitem{Wong_2020}
P.-H. Wong.
\newblock Democratizing algorithmic fairness.
\newblock {\em Philosophy \& Technology}, 33(2):225--244, 2020.

\bibitem{Wong_2012}
Y.~Wong, M.~T. Harandi, C.~Sanderson, and B.~C. Lovell.
\newblock On robust biometric identity verification via sparse encoding of
  faces: Holistic vs local approaches.
\newblock In {\em International Joint Conference on Neural Networks (IJCNN)},
  2012.

\bibitem{woodward_our_2020}
E.~Woodward, R.~Hill, P.~Harkness, and R.~Archer.
\newblock {\em Our {Knowledge} {Our} {Way} in Caring for {Country}}.
\newblock {NAILSMA} and {CSIRO}, 2020.
\newblock Online. {URL:} {http://www.csiro.au/ourknowledgeourway} (accessed: 07
  Sep 2022).

\bibitem{QiangYang_2019}
Q.~Yang, Y.~Liu, T.~Chen, and Y.~Tong.
\newblock Federated machine learning.
\newblock {\em ACM Trans.~Intelligent Systems and Technology}, 10(2):1--19,
  2019.

\bibitem{Agerfalk_2022}
P.~J. Ågerfalk, K.~Conboy, K.~Crowston, J.~E. Lundström, S.~L. Jarvenpaa,
  S.~Ram, and P.~Mikalef.
\newblock Artificial intelligence in information systems: State of the art and
  research roadmap.
\newblock {\em Communications of the Association for Information Systems},
  50(1):420--438, 2022.

\end{thebibliography}

\end{document}